# Interference Simulator for the Whole HF Band: Application to CW-Morse

E. Mendieta-Otero, I. A. Pérez-Álvarez, *Member*, *IEEE*, and B. Pérez-Díaz

*Abstract*—In this paper, we use jointly a model of narrow band interference and a congestion model to model and implement an interference simulator for the whole HF band. The result is a model to generate interfering signals that could be found in a given frequency allocation, at a given time (past, present or future) and for a given location. Our model does not require measurements and it is characterized by its ease of use and the freedom it offers to choose scene (modulation, location, week, year, etc.). In addition, we have defined a generic modulating function and the conditions to model a "contact" CW (Continuous Wave)-Morse, who meets the usual standards of contest. Consequently, our interference model in conjunction with the CW-Morse modulating function designed, it results in a specific CW-Morse model for amateur contests. As an example of the simulation model, we simulate the CW-Morse communications on the contest "ARRL Field Day 2011".

## I. INTRODUCTION

The present needs for more accurate modeling and simulation of the HF band (formally, 3 – 30 MHz [1]) have motivated the development of wideband models and their implementation in wideband channel simulators. For example, there is a need for modeling the HF channel bandwidth to assist in the development of systems based on the recently promulgated US military standard MIL-STD-188-110C that contains an appendix (Appendix D) defining a new family of wideband HF data waveforms. Moreover, fully realizing the potential of these new waveforms will require enhanced capabilities in other elements of HF communications systems such as Automatic Link Establishment (ALE) systems.

While propagation conditions have been studied extensively over several decades, there are relatively fewer published studies that have looked at noise and interference in the HF band. Moreover, it is a well known fact that interference from legitimate users is one of the most common problems encountered in the use of the HF spectrum. For the radio amateur service several frequency bands are allocated [2] and [3: 982], and properly licensed stations are permitted to operate at any frequency within these bands. It is remarkable that in the 2000's there were about three million amateur radio operators worldwide [4]. Given the global scope and the rules of participation [5] of major radio amateur contests (ARRL Field Day, ARRL International DX Contest, IARU HF Championship, CQ WorldWide Contest, etc.), it is reasonable to expect:

1) a high number of contestants;
2) at reception, a wide range of values for: the powers, the frequencies, the start times of communications, etc.

Thus, the major radio amateur contests can be a good scene to evaluate a communications system subjected to a high degree of "interferences". Also, interference from other users is frequently more important than the man-made noise [6] from incidental radiators or atmospheric noise from lightning [7].

Therefore, for the evaluation of HF equipment in general, and wideband systems in particular, comes the need for interference simulators throughout the HF band. If in addition, the simulator is capable of generating interference in a temporary situation or in any location, the simulations for the situation in the past could serve as a benchmark to the own simulator in its most desirable application: generating interference in a future environment. With this objective, we use jointly a model of narrow band interference and a congestion model to generate a new model and implement an interference simulator for the whole HF band. The congestion value indicates the probability of interference under the given conditions [3].

Communication methods of radio amateurs are diverse. However, we have focused our attention on CW-Morse. Its principle of operation is based on the carrier-no carrier signal, which means a great simplicity in the hardware implementation, essential in emergency situations. With regard to its ability to interfere, CW is a signal that focuses all its energy in a very narrow bandwidth. However, our simulator opens the possibility of implementing any type of modulation. The "ARRL Field Day 2011", organized by the American Radio Relay League (ARRL), is used as application example of the simulator.

As previous models in the same line that ours, we have found [8], [9] and [10]. The main difference of the model in [8], [9] with ours is that in our model the amplitudes of interference are deduced from a model of congestion [3] and in

Manuscript received September 10th, 2012. This work was supported by the Ministerio de Economía y Competitividad under grant TEC2010-21217-C02-01.
E. Mendieta-Otero, I. A. Pérez-Álvarez and B. Pérez-Díaz are with the Institute for Technological Development and Innovation in Communications (IDeTIC), Las Palmas de Gran Canaria, Spain (e-mail: emendieta@idetic.eu).
E. Mendieta-Otero and I. A. Pérez-Álvarez, are with the Department of Signals and Communications, Las Palmas de Gran Canaria University (ULPGC), Spain.
Digital Object Identifier

[8] and [9] are deduced from a model developed by Hall [11]. Basically, the congestion model election has been made to have the capability of extrapolation in time in our model and in our simulator. Our model and the model in [10] use a model of congestion to deduce the amplitudes of the interference. However, the main differences of our model with that of [10] are: 1) the model of [10] does not include modulation; 2) in [10] the number of interferences and the frequencies of the interference, are fixed: 800 sine waves 250 Hz apart, giving a total bandwidth of 200 kHz. In our model, the number and frequency of interference are Random Variables (RVs) limited by the event one wishes to simulate.

This paper is organized as follows. Section II describes a model of narrow band interference that is the fundamental structure of our model of interference: a sum of sinusoids. Section III describes our interference model for the whole HF band (1,606 – 30 MHz). This is the frequency range of our model of interference since it is the frequency range of the congestion model that we used in our model. Section IV describes a model of congestion from which we deduce, in our model, the values of amplitude of the interference. Section V: 1) clarifies how these amplitudes are deduced; 2) develops the modulation chosen to the interference; 3) presents the application example chosen, the "ARRL Field Day 2011"; 4) describes the temporal control processes (onset, duration, etc.) of interference of our model. Section VI describes the implementation of our simulator based on our model. Section VII exposes the simulation results for the selected example of application, the "ARRL Field Day 2011", at a given time and place. Section VIII presents future work. Section IX concludes the paper.

## II. NARROW BAND INTERFERENCE MODEL

Our model of interference, which is discussed in Section III, has the same fundamental structure as that of a narrow band model. Thus, the fundamental structure of both models of interference, our model and the narrow band model, is a sum of sinusoids. In the narrow band interference model chosen as a starting point [8] and [9], the narrowband interferers can be written as

$$i(t) = \sum_l C_l \cdot e^{j(2\pi \cdot \Delta f_l \cdot t + \phi_l)} \quad (1)$$

where $l$ is the number of the interference, $C_l$ are the amplitudes of the sine waves, $\Delta f_l$ are the baseband frequencies of the sine waves ($\Delta f_l = f_l - f_0$), $f_l$ is the carrier frequency of the interference (RF), $f_0$ is the offset frequency to baseband and $\phi_l$ are random phases. The narrow band interference model is based on the results from several case studies [8] and [9]. The data consisted of 42 one-second records of the digitized baseband signal [8]. The interference model development involved examining statistical characteristics of measured interference and developing a model of the interference waveform that exhibits those same characteristics [8].

In (1), the narrow band reference model, the probability density function (pdf) of the amplitudes $C_l$ of the interferers is modeled by the amplitude pdf of the model developed by Hall [11]

$$p_C(C) = (\theta_C - 1) \cdot \gamma_C^{(\theta_C - 1)} \cdot \frac{C}{\left(C^2 + \gamma_C^2\right)^{(\theta_C + 1)/2}} \quad (2)$$

where $C_l$ depends on two free parameters, $\theta_C$ and $\gamma_C$ [8] and [9]. $\theta_C$ and $\gamma_C$ are chosen to fit a reference pdf of the amplitudes of the interference measures.

Summarizing, the narrow band model (1) and (2):
1) depends on the fit of the parameters $\theta_C$ and $\gamma_C$ to specific measurements;
2) therefore depends on the completion of measures in the geographical and temporal environment that you want to simulate;
3) the number of interference, $l$, is fixed to get the best fit to the measurements and also the start times and duration are chosen to get the best fit to the measurements [8], [9];
4) does not explicitly include any modulation.

Our model, which is presented in the following section, is intended to overcome these disadvantages such that:
1) it does not require measures nor the adjustment of free parameters to those measures;
2) it allows the simulation of future scenarios;
3) it is not restricted to a fixed number of interferences and it specify the start and duration of each interference;
4) it includes a modulating function.

The major objectives 1) and 2) will be treated in sections III, IV and paragraph *A. Amplitude* of Section V. To this end, the $C_l$ of (1) is replaced by $I_l$, a RV deduced from a model of congestion. This deduction is further developed in paragraph *A. Amplitude* of Section V.

Objective 3), the number of interferences and their temporal evolution, will be discussed in paragraph *C. Initial Time and Duration* of Section V.

Objective 4), include a modulating function, will be discussed in Section III and in paragraph *B. Modulation* of Section V.

## III. INTERFERENCE MODEL FOR THE WHOLE HF BAND

As mentioned, the idea of interference as the sum of sinusoids is taken from a narrow band interference model [8] and [9]. In our model, the waveform of the whole HF band (1,606 – 30 MHz) interference is represented as a voltage of the form

$$v(t) = \sum_l I_l \cdot m_l(t) \cdot \cos[2\pi \cdot F_l(t) \cdot t + \Phi_l(t)] \quad (3)$$

$$\text{where } F_l(t) = f_l + \Omega_l(t) \quad (4)$$

$$\Phi_l(t) = \phi_l + \theta_l(t) \quad (5)$$

where $l$ ($l$ = 1, 2, …) is the number of the interference, $I_l$ is the amplitude, $m_l(t)$ is an amplitude modulation function, and $f_l$ and $\phi_l$ retain the same meaning as in (1). $\Omega_l(t)$ and $\theta_l(t)$ are frequency and phase modulation functions, respectively. For example, for an M-ary PSK, (3) takes the form

$$v(t) = \sum_l I_l \cdot \cos[2\pi \cdot f_l \cdot t + \Phi_l(t)] \qquad (6)$$

where $f_l$ and $\phi_l$ retain the same meaning as in (1) and $\theta_l(t)$ represents each of the $M$ possible values of phase, depending on the modulating signal: $(2\cdot\pi\cdot n)/M$ where $n$ = 0,…, $M$-1. For our application, CW-Morse, (3) takes the form

$$v(t) = \sum_l I_l \cdot m_l(t) \cdot \cos(2\pi \cdot f_l \cdot t + \phi_l) \qquad (7)$$

where $f_l$ and $\phi_l$ retain the same meaning as in (1) and $m_l(t)$ will be deducted in Section V.

In (4), as in (1), the frequencies $f_l$ are uniformly distributed [8] and [9]. In (5), as in (1), the phases $\phi_l$ are uniformly distributed between 0 and 2·π [8], [9] and [12]. In our model, (3), the operating frequencies must be within the operating bands of congestion model used (1,606 – 30 MHz), which is discussed in the next section. That is, the random nature of the frequencies, $f_l$, and phases, $\phi_l$, is the same as the reference narrow band model, adjusting the frequency range in which we are interested.

As mentioned, the amplitudes of the interference, $I_l$ in (3), (6) and (7), are derived from a model of congestion. Congestion is defined as the probability that a randomly selected channel within a given spectrum allocation will exceed a specified power threshold [3]. As noted, the conditions of validity of the model of congestion impose the range of possible values of the random frequency, $f_l$ in (4). The amplitude modulation function, $m_l$, depends on the type of interference you want to simulate, that in the case of this paper is CW. The simulator implemented, generates the band-pass signal (RF) given by (3), as well as its baseband (IQ) version, as discussed in Section VI.

As will be discussed in the next section, the value of the amplitude of each interference, $I_l$, depends, among other parameters, on the geographical location of the receiver and the instant of reception. These amplitude values, $I_l$, are based on weekly values centered at noon or midnight. Therefore, our model is valid for weekly periods, with the highest accuracy around noon and midnight.

IV. CONGESTION MODEL

As previously stated, by using a congestion model we deduce the values of the amplitudes of the interference, $I_l$, in (3). There are other studies about HF interference modeling but none of them has been so thoroughly verified by measurements as the congestion model developed at the University of Manchester Institute of Science and Technology (UMIST) [13].

We used the model of congestion of [3] in the development of our model of interference. Although more recent congestion models have been proposed, the one we have chosen is that which offers the highest transparency and utility to our application. Thus, the model chosen [3]:

1) is based on two functions easy to implement in, for example, Matlab;
2) specifies all the values of the constants that are part of these functions;
3) offers the experience of measures and models developed from the seventies. In more recent studies it is possible to predict the congestion corresponding to the hour of the day and the day of the year [14] and [15]. For this case, we think the use of neural networks means less transparency and simplicity that the use of simple and well-defined functions.

Congestion, $Q_k$ (which is a probability), varies with field-strength threshold level, frequency allocation [2] and [3], time, bandwidth, location and sunspot number. In the UMIST model, and therefore in our model, the HF spectrum is divided in $k$ frequency allocations. This congestion model is based on weekly measurements, in Northern Europe, using a calibrated low-angle monopole antenna [3].

If $Q_k$ is the modelled value of congestion for the $k$th frequency allocation ($k$ = 1, 2, ..., 95), we have [3]

$$Q_k = \frac{1}{1+e^{-y_k}}. \qquad (8)$$

The model estimates values of congestion $Q_k$, in the range 0 to 1. In this case, the model index functions, $y_k$, apply to stable-day and stable-night ionospheric conditions. These conditions correspond to two periods, each with a duration of about three hours, centered on local midday and local midnight. These index functions can be expressed as

$$y_k = \alpha_k + B_k \cdot E \qquad (9)$$

where $\quad \alpha_k = f(f_k, BW, SSN, \theta_w, \theta_{long}, \theta_{lat}) \qquad (10)$

$$B_k = B_0 + B_1 \cdot f_k + B_2 \cdot f_k^2 \qquad (11)$$

where $E$ is a certain field-strength threshold in dBV/m, $B_0$, $B_1$, and $B_2$ are constants, $f_k$ is the central frequency of allocation $k$ and $\alpha_k$ is a function of all other parameters of the congestion model, except the threshold $E$ [3:983]. In the function (10) there are 142 estimated model coefficients for the stable-day index function, from A$_1$ to J$_{3,2}$ [3:984], and there are 139 estimated model coefficients for the stable-night index function [3:985]. Namely, for each of the two cases, stable-day and stable-night, $\alpha_k$ for the $k$th frequency allocation depends on the following variables: $f_k$, the receiver filter bandwidth $BW$, the SunSpot Number $SSN$, the $week$ of the year defined through $\theta_w$, and longitude and latitude represented by $\theta_{long}$ and $\theta_{lat}$, respectively. In $\alpha_k$, the *week* variable, on which depends $\theta_w$, varies in range 1-52 and the *SSN* variable has monthly values [3].

In our simulator, to each RV $f_l$ of (4) is assigned a value $k$ of frequency allocation ($k = 1, …, 95$; that is, the permitted values of $f_l$ are between 1,606 MHz to 30 MHz [3: 982]). With this value of $k$ is calculated the central frecuency of allocation $k$: $f_k$ of (10) and (11). As will be discussed in Section V, the amplitude, $I_l$ of (3), of each interference $l$ ($l = 1, 2, …$) with carrier frequency $f_l$ (4) is calculated from the values of $\alpha_k$ of (10) and $B_k$ of (11).

The ability of the congestion model to estimate future congestion values and to extrapolate in space (over the measurement sites) was investigated in [3], and it was concluded that such capabilities were clearly demonstrated.

### A. SunSpot Number Data

As already mentioned, the variable $\alpha_k$ in (9) and (10) depends on *SSN*. Thus, the ability to predict the *SSN* values will affect the quality of the model of congestion and, consequently, the quality of our model.

In our simulator and in [3], the actual and the past *SSN* values are based on the Monthly International Values, published by the World Data Centre for Solar-Terrestrial Physics (WDC) [16], see Fig. 1.

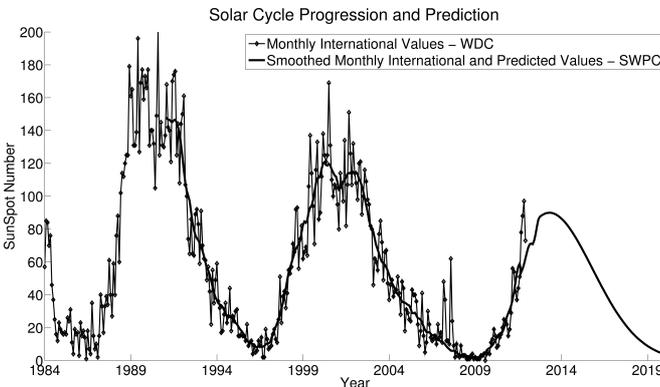

Fig. 1. Solar progression and prediction for the years 1984 to 2019 from the WDC and the SWPC.

Also, the actual and predicted SSNs may be obtained from the NOAA's Space Weather Prediction Center (SWPC-NOAA) [17]. In our work, for future SSN values we employ the smoothed monthly predicted values from the SWPC-NOAA, from 2012 onwards. Fig. 1 depicts the progression and prediction of the Solar Cycle for the years 1984-2019.

## V. DEVELOPMENT OF THE INTERFERENCE MODEL FOR THE WHOLE HF BAND

### A. Amplitude

As aforementioned, the amplitudes of the interferences of our model are derived from a model of congestion. The first step is to obtain the electric field received. To obtain it, we combine (8) and (9) of the model of congestion, with the fact that the congestion is defined as a probability (of exceeding a threshold) and hence gives the cumulative probability (of being below a threshold)

$$\text{Prob}(E) = 1 - Q_k = \frac{1}{1+e^{\alpha_k + B_k \cdot E}} \quad (12)$$

inverting the result we obtain

$$E(\text{Prob}) = \frac{\ln\left(\frac{1-\text{Prob}}{\text{Prob}}\right) - \alpha_k}{B_k} \quad (13)$$

and treating the cumulative probability Prob as a RV uniformly distributed between 0 and 1, we obtain the RV electric field $E$ in dBV/m [18]. The next step is the conversion of the electric field into power.

In summary, the amplitudes of the individual interferences in our model, $I_l$ in (3), are those corresponding to the average powers ($P_l$). These average powers depend on the RVs $E$. If Antenna Factor ($AF$) is defined as the ratio of the incident electromagnetic field to the output voltage from the antenna, it can be deduced the relationship between $P_l$ and $E$:

$$P_l = 20 \cdot \log_{10}\left(E/10^{-6}\right) - AF - 10 \cdot \log_{10}(R) - 90 \quad (14)$$

where $P_l$ is the average power in dBm delivered by the antenna terminal (or antenna amplifier) to a load of $R$ ohms, $AF$ is in dB/m, and $E$ is in V/m. $AF = 10$ dB/m in [3].

The electric field $E$ is a function of both $\alpha_k$ and $B_k$ according to (13). Therefore, as $\alpha_k$ and $B_k$ are parameters of the congestion model, the amplitudes of the individual interferences, $I_l$ in (3), are functions of the same variables that the congestion model: $f_k$, $BW$, $SSN$, $\theta_w$, $\theta_{long}$, $\theta_{lat}$, *etc.* (see Section IV).

### B. Modulation

In our model, the modulation type to assign each frequency allocation has to be selected by the user (along with the date and location that you want to simulate). The reasons for this manual selection are: 1) greater simplicity in implementation; 2) in some amateur bands can coexist various types of modulations; 3) frequency allocations are not always respected; 4) the user has greater freedom of simulation.

In this section: 1) the choice of modulation CW-Morse is justified; 2) the communications event that has served as a reference for the implementation of our model, namely the "ARRL Field Day 2011", is discussed; and, 3) our use of the keyword "PARIS" for CW-Morse, is justified. All this leads to the expression of the modulating function, $m_l(t)$, presented at the end.

We chose the modulation CW-Morse, because of:
1) simplicity of software and hardware implementations;
2) it involves narrow band signals such as the narrow band model of reference (1);
3) the transmitter power is concentrated on a single carrier with ASK (Amplitude Shift Keying) modulation and with a bandwidth much less than the other available modulations in HF; consequently, the CW-Morse emissions are perceived as high-power interference with respect to other emissions; and
4) it is easier to define a generic modulating function.

When you want to include a specific modulating function in (3), (4) and (5), such that involves ease of use of the simulator, the question that may arise is: what modulating signal exhibits the common features to many (or infinite) modulating signals for a given communication environment? For example, what could be the standard modulating signal for: a SSB (Single Side Band) contact on amateur radio, a CW-Morse contact on amateur radio, an AM broadcasting, a M-ary PSK military communication, etc. A contact, or exchange of information between two amateur radio stations, is often referred to by the Q code as a QSO [19]. Given the recommendations relating to Morse code [20] and to CW-Morse contacts [19] and the contests rules [5]: the CW-Morse contact is easier to analyze than the other types of communications.

As an example of application of the simulator we chose an amateur radio contest, the "ARRL Field Day 2011" organized by the ARRL, because among other reasons (see *C. Initial Time and Duration*) in the amateur bands the randomness of emissions is higher than in the fixed service bands or broadcast services. The "ARRL Field Day" is always the full fourth weekend in June, beginning from 1800 UTC Saturday and ending at 2100 UTC Sunday [5].

In the "ARRL Field Day 2011" there were 577,181 CW QSOs or "Continuous Wave contacts" [5]. The average number of CW "contacts" in the "ARRL Field Day", from 2005 to 2011, was 530,550. It is often limited to a minimum exchange of identification of stations (IDs). Thus, each QSO involves a total of five transmissions, three by the requesting "contact" and two from the accepting one [19], [20], and [5].

As the five "contacts" transmissions take place on a common frequency and with the greatest possible continuity over time, in our simulator, each QSO will be considered as a single communication. Thus we conclude that in the "ARRL Field Day 2011" the average was 6.68 communications per second for 24 hours distributed among the amateur bands according to the contest rules [5].

The Morse code speed is measured in Words Per Minute (WPM). There is a standard word to measure operator transmission speed: "PARIS". The 50 bit keyword "PARIS" [20] is used as modulation pattern in our simulator. The keyword "PARIS" contains a percentage of non-signal of 56%, similar to a typical QSO with an 45%.

Fig. 2 (a) shows a CW-Morse signal corresponding to the keyword "PARIS". If we zoom into a portion of the signal (b), you can see the signal corresponding to the "dot", i.e. the duration of the presence of carrier signal. The duration of this "dot" is the result of a Morse code speed of 60 words per minute, according to the known relationship given by:

$$WPM = 1.2 / dot \qquad (15)$$

being *WPM* speed in words per minute (using the keyword "PARIS") and *dot* the length of the "dot" in seconds. The actual range for manual Morse code speed is usually between 10 and 75.2 WPM. In our simulator, the Morse code speed is a RV since it depends on another RV, namely the *dot*.

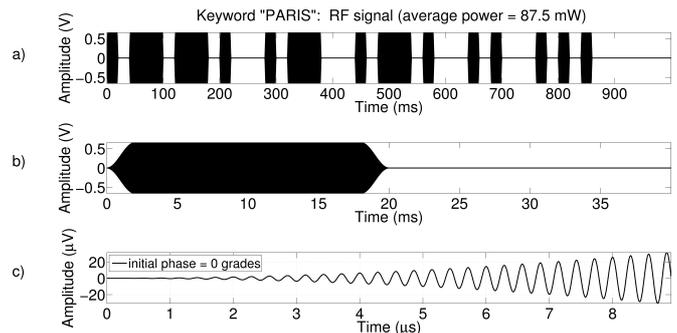

Fig. 2. a) Simulation of a CW-Morse signal corresponding to the keyword "PARIS" for a 3.5 MHz carrier; b) enlargement of a portion of the signal with a "dot"; c) part of the "rise" in amplitude of a "dot".

In summary, the modulating function $m_l(t)$ in (7) for our application of CW-Morse can be expressed as

$$m_l(t) = \begin{cases} PARIS_l(t), & t_{li} \le t \le (t_{li} + t_{ld}) \\ 0, & \text{other } t \end{cases} \qquad (16)$$

where $t_{li}$ is the initial time, $t_{ld}$ is the duration of the interference, QSO or communication (onward, interference) and $PARIS_l(t)$ is a unitary envelope corresponding to the keyword "PARIS" repeated indefinitely for the duration $t_{ld}$ of the interference number *l*. The envelope generated by (16) is of unit value.

### C. Initial Time and Duration

We use a Poisson process (with parameter $\lambda$ equal to the expected number of interferences that occur per unit time) for the number of interfering signals in a time interval which occur continuously and independently of one another. This is the more frequently encountered situation, where there may be many (10, 20, …,∞) potentially interfering sources [12]. Consequently, the probability distribution of the waiting time until the next interference, $\Delta t_n$, is an exponential distribution (with parameter $\mu = 1/\lambda$ equal to the expected interarrival time). Thus, the initial times of interference, $t_{li}$, are given by

$$t_{li} = \sum_{n=0}^{l-1} \Delta t_n \qquad (17)$$

where $l = 1,2,\ldots$, and $t_{0i} = 0$. It is also assumed that the durations of the interference, $t_{ld}$, are exponentially distributed independently among them [21]. Table I shows the statistical parameters chosen for interference in the "ARRL Field Day 2011".

The criteria used for the other temporary variables are:
- each interference, $l$, consists of 331 "dots", ie

$$dot_l = t_{ld} / 331; \qquad (18)$$

- rise times, $T_{lr}$, and fall times, $T_{lf}$, of each envelope are identical and equal to

$$T_{lr} = T_{lf} = dot_l / 10; \qquad (19)$$

- as a consequence of the above, for each interference, $l$, we have a random value of: $t_{li}$, $t_{ld}$, $dot_l$, $WPM$, $T_{lr}$ and $T_{lf}$.

The justification of (18) is based on the typical QSO or interference seen in *B. Modulation*. Thus we establish the total communication time is $331 \cdot dot$, assuming that $dot$ duration is the basic unit of time measurement. For example, if the RV duration to the interference $l$, $t_{ld}$, is 10 seconds, according to (18) the variable $dot_l$ is 30 ms and according to (15) the variable $WPM_l$ is then 40 words per minute.

The rise and fall of the envelope of the carrier of each interference has a raised cosine shape and its duration has been chosen as (19), as seen in Fig. 2 (b) and (c). These shape and duration are common in radio for: minimum bandwidth, minimal generation of spurious and to increase the average life of hardware power amplifiers. In our simulator, the values of the RV rise time, $T_{lr}$, and fall time, $T_{lf}$, of the "dot" are constant over each interference $l$.

Given the usefulness of the keyword "PARIS", our simulator sequentially sends this keyword until the end of each interference. Please, recall that each interference begins at $t_{li}$, and lasts $t_{ld}$.

Hence, the total number of interference $N$ (such that $l = 1, 2, ..., N$), is a RV which depends on:
1) the expected number of interferences that occur per unit of time;
2) the total time you want or can simulate;
3) and, of course, depends on the resources available in order to simulate: memory availability, computational load manageable, etc.

As noted, to deduce the number of interferences and the waiting times until next interference we must know the parameter λ (the expected number of interferences that occur per unit time). Obviously, to find the value of λ we can either measure it or go to a reliable source of information. The radio amateur contests are usually a quick and simple source of knowing the number of communications that have been established in a given time. Some of the radio amateur contests with more participation are: ARRL Field Day, ARRL International DX Contest, IARU HF Championship and CQ WorldWide Contest. The ARRL is a good example of ease in obtaining information concerning amateur radio contests [5].

Summarizing, the key differences between our model (3) and the narrow band model (1) are:

1) the narrow band model gives more freedom to fit the simulation to the measurement; for this purpose, the user of the simulator must adjust: the number of interferences, the pdf of the amplitudes of the simulated interferences to get the best fit to the measurements, etc.;
2) by contrast, our model is much more friendly and dynamic: our model allows freedom and ease in selecting the date and / or location of the simulation; and the number, duration and the amplitudes of the interferences are chosen automatically by the simulator, etc;
3) there have been no comparisons between our model and the narrow band model for deciding which of the two models best fits reality;
4) the narrow band model requires measurements and our model does not;
5) our model simulates future interferences.

VI. IMPLEMENTATION OF THE INTERFERENCE SIMULATOR

This section summarizes the implementation of the simulator for our model of interference for the whole HF band. The basic equation describing the model is (3) and its parameters have been described in previous sections. In the example that will be presented in this section, the simulator output signal is at the input of a receiver under the following conditions:
- Operating band: HF amateur frequency allocation [2] and [3]
- Location: Las Palmas de G.C., Spain, that is, latitude 28º and longitude -15.35 º
- Date: fourth full weekend in June, 2011, about three hours, centered on local midday (1300 UTC)
- Modulation: CW-Morse

As discussed in the section *B.Modulation*, in the described implementation, our simulator deals only with the amateur bands because of their high randomness. Thus, of the 95 possible frequency allocations, in the following example we only consider the amateur frequency allocations, i.e. the allocations: 11, 26, 37, 50, 62, 71, 82, 92, 93 and 94. In any case, our simulator only chooses values for the RVs carrier frequencies, $f_l$ in (4), who are lying within the 95 frequency assignments considered in the model of congestion of reference [3]. The simulator has been implemented with Matlab. Table I shows a summary of the parameters of the simulator.

Fig. 3 shows the pdf of the powers of interference, $P_l$ in dBm, whose values are obtained by (13) and (14). This pdf is symmetrical around $-\alpha_k/B_k$, thus the expected value $E[P_l] = -\alpha_k/B_k$ (dBV/m) and the value of $B_k$ decides the spread of $P_l$ [18]. Then, according to (11), the higher the frequency, the greater the spread of the amplitudes of the interference, $I_l$. For the values of $\alpha_k$ and $B_k$ of the example in Fig. 3 we have $E[P_l]$ = -107.3391 dBm. Given $B_k$, the value of $\alpha_k$ determines the average value of $P_l$ (dBm). Typical values of $\alpha_k$ are between -16 and -9 [18] and, for typical values of $B_k$ about -0.1, the resulting average values of $P_l$ lie between -160 and -90 dBm. Therefore, these are the ranges of average power of interference we can expect in our simulator.

TABLE I
STATISTICAL PARAMETERS OF THE SIMULATOR

| Parameters | Distribution | Statistic |
|---|---|---|
| carrier frequency ($f_l$) | uniform | values within the amateur frequency allocations |
| phase ($\phi_l$) | uniform | $(0 - 2\pi)$ |
| waiting time ($\Delta t_n$) | exponential | $\mu = 1/\lambda$; $\lambda = 6.68$ interference/s |
| duration of each communication ($t_{ld}$) | exponential | $\mu_d = 10$ s |
| envelope amplitude ($I_l$) | depending on Congestion | $\alpha_k, B_k$ |
| envelope modulation ($m_l$) | Modulation CW-Morse with the keyword "PARIS" repeated until the end of each interference | |

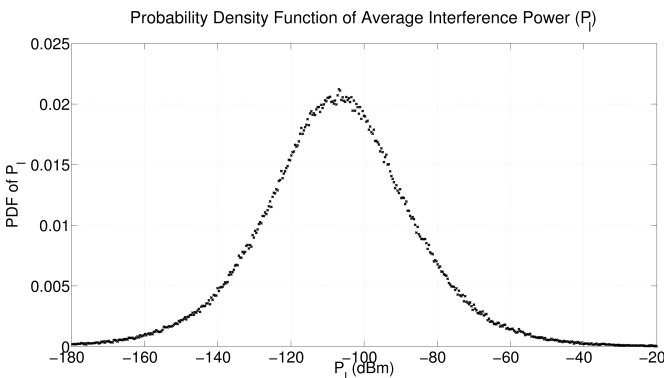

Fig. 3. Pdf of the RVs powers of interference with $B_k$ = -0.084873, $\alpha_k$ = -10.8077, stable day and conditions of location, date, etc. of this section.

In short, the simulator produces each interference under the random parameters already discussed and also combines in time any interference according to its initial time and duration. Fig. 4 shows the first 16 seconds of the output signal of the simulator.

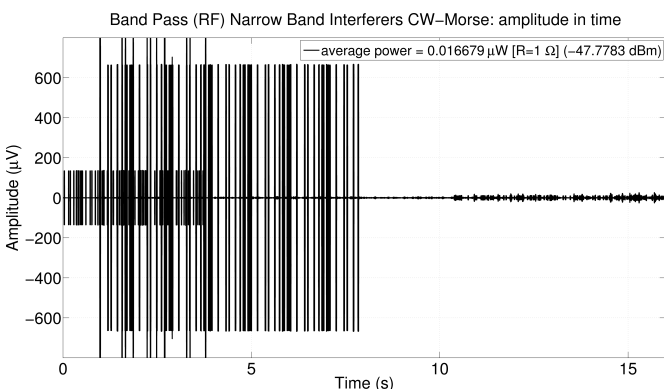

Fig. 4. Simulated total interfering signal.

Fig. 5 has a greater detail of the evolution both in time and frequency of the simulated total interfering signal. The signal shown in Fig. 5 illustrates the temporal coincidence of several interference.

With the objective of reducing both the computational load and memory requirements, the simulated total interfering signal is subjected to a digital down conversion (DDC), moving the center frequency of the HF band (or any other frequency that the user of the simulator want to choose between 1,606 and 30 MHz) to 0 Hz. The base-band complex signal is further decimated by a factor of 2, and then filtered.

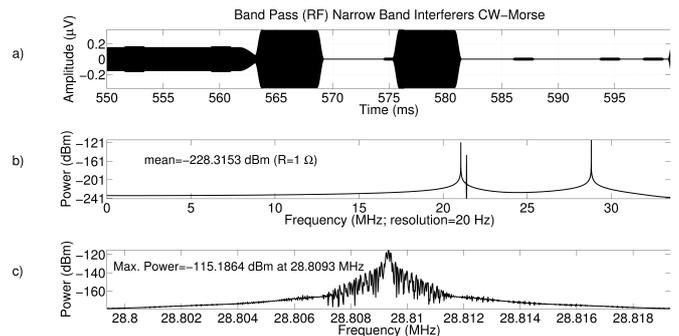

Fig. 5. Evolution in time and frequency of the simulated signal: a) from 550 to 600 ms, b) its spectrum, c) zoom of the spectrum.

VII. SIMULATION RESULTS

This section will present some statistical measures performed on the simulated total interfering signal from the previous section. Its purpose is confirm the random characteristics imposed on the simulated signal and also its possible future comparison to actual measurements statistics in an "ARRL Field Day". In Fig. 6, we can observe the evolution in time and frequency of the simulated total interfering signal during 0.5 seconds.

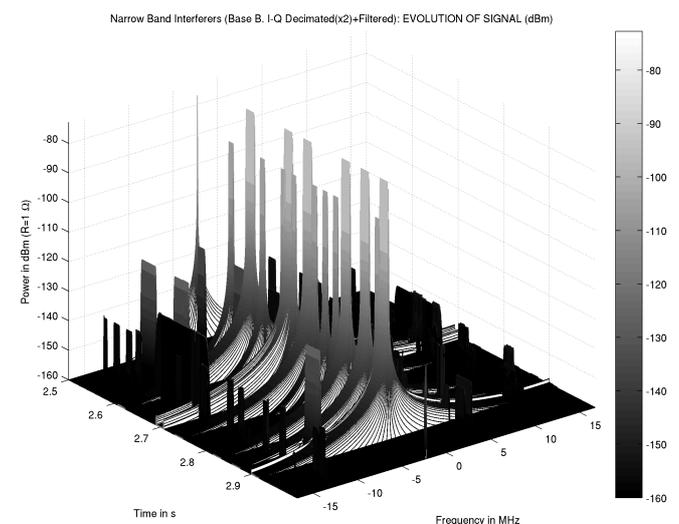

Fig. 6. Time evolution of signal spectrum for the simulated total interfering signal. The frequency resolution is 8192 Hz. For this time interval the maximum power was found to be -72.7116 dBm at a frequency of -2.2856 MHz (14.2144 MHz in RF).

In Fig. 6, we can see some important characteristics:
1) possibility of large differences of power between the various interference as expected according to their power pdf (see also Fig. 3);
2) some degree of continuity in the time domain if all individual interferences are superimposed, even though the evolution of Morse code for the keyword "PARIS" in individual interference shows a remarkable absence of carrier (56%), see also Figs. 2 and 5;
3) impulsivity in the frequency domain, although the carrier frequencies are constant for each interference these spectral components are absent for about 56% of the duration of each interference;
4) the width (time duration) of each "dot" remains constant in each interference because it is assumed that for each communication both interlocutors have the same Morse code speed and, therefore, the same width of "dot".

### A. Amplitude Probability Distribution

The Amplitude Probability Distribution (APD) succinctly express the probability that a signal amplitude exceeds a threshold. The APD provides information about an interference signal to estimate the performance degradation of an victim receiver [22].

In this paper APDs are plotted on a Rayleigh probability graph [22]. In this paper the axes of the representation of APD represent the power of the decimated I-Q interference signal in dB above $k \cdot T_0 \cdot B$ versus the percent-of-time the power is exceeded. The power $k \cdot T_0 \cdot B$ is the thermal noise present in every receiver. In this case, the bandwidth chosen for the thermal noise is 100 Hz which coincides with the IF filter of some commercial receivers.

In Fig. 7, we can see the curve of the normalized APD for three signals in the time domain: 1) the simulated total interfering signal; 2) a sinusoid (peak amplitude = 400 µV); and 3) an "impulsive" signal (a signal with a high ratio of peak amplitude to the root mean square amplitude) formed by: 2 µs without carrier signal + 1 µs with carrier signal (including rise and fall times of envelopes equal to 0.1 µs and peak amplitude = 400 µV ) + 3 µs without carrier signal + 1 µs with carrier signal + without carrier signal until 16 seconds, the total duration of the signal. The APD of a sinusoid signal, in the time domain, is a flat line from the lowest to certain percentile on a Rayleigh graph [22]. The powers of the impulsive signals are concentrated in short periods of time separated in time by quiescent intervals, which results in curves similar to that of Fig. 7. According to Fig. 7, the step shape (or flat line from the lowest to certain percentile) of the normalized APD of the interference, in the time domain, implies a shape similar to the sinusoidal. Although individual interferences contain an 56% average percentage of non-signal, this result is expected because there are many individual interferences that overlap in time, see Figs. 4, 5 and 6.

In Fig. 8, we can see the curve of the normalized APD of the three signals mentioned, in the frequency domain. Comparing Figs. 7 and 8, we see that the APD of the sinusoid in the time domain, is very different from the APD of the sinusoid in the frequency domain. The same happens with the "impulsive" signal. This is expected because the bandwidths associated with each of these signals (very large in the case of the "impulsive" signal).

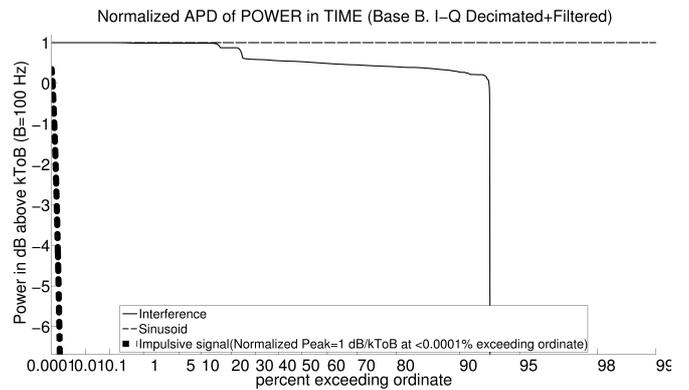

Fig. 7. Normalized APD for: the simulated total interfering signal, a sinusoid and an "impulsive" signal, in the time domain.

Both Figs. 7 and 8 indicate an interference APD closest to the APD for the sinusoid than for the "impulsive" signal. As shown in Figure 6, the impulsive behavior seems higher in the frequency domain than in the time domain since in the frequency domain there is no overlap between individual interferences. Namely, this higher impulsiveness in the frequency domain is due to the fact that frequencies of the individual interferences are relatively far from one another, and also to that the individual interferences are present only for short periods of time in which there is carrier. However, in Figs. 7 and 8 the APD of the total interfering signal not clearly shows that the impulsiveness is higher in the frequency domain than in the time domain. This is because the APD is not a good indicator of the arrival times of the signal amplitudes, as in the case of the Level Crossing Statistics, some of which are discussed in the following section.

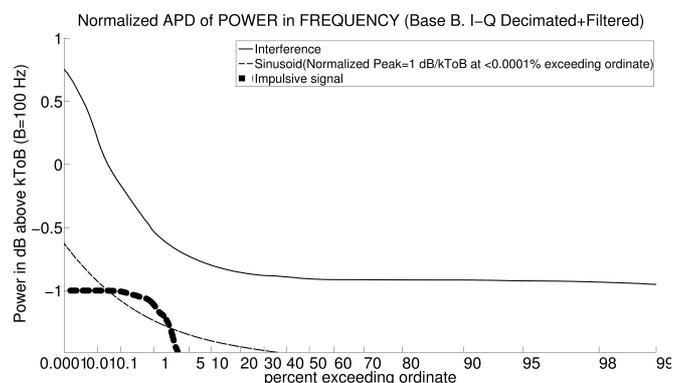

Fig. 8. Normalized APD for: the simulated total interfering signal, a sinusoid and an "impulsive" signal, in the frequency domain.

*B. Level Crossing Statistics*

Computation of BERs in many modern digital receivers requires statistics describing the time of arrival of signal amplitudes [22]. Moreover, due to the nature of our simulated signal, there may be a large amount of carrier absence, about 56% (see Fig. 2), or a large number of overlapping small amplitude values (see Figs. 4, 5 and 6). This would result in pdfs for the envelopes of voltage and power with narrow and sharp peaks due to the values equal to zero. These pdfs may mask the distributions present in the signal. Using level crossings statistics we can overcome this inconvenience by setting a minimum threshold of zero. In this way, we only take into account how much the values exceed this threshold. This type of situation is common not only for CW-Morse signals but for pulsed signals and impulsive noise.

To study the envelope values we have in the simulated signal, we used the Level Crossing Distribution (LCD). The LCD is the distribution of the number of upgoing crossings through some levels or thresholds of the signal envelope. Fig. 9 shows a representation of the normalized LCD for the three signals discussed in the previous section, and it can be appreciated the most common envelope levels in the interference (see also Fig. 4, 5 and 6) as well as its great variability (a high value of Max. LCD) with respect to the sinusoid and the "impulsive" signal.

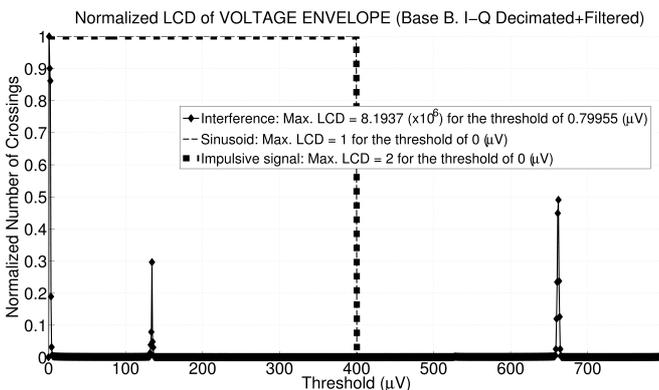

Fig. 9. Normalized LCD of the voltage envelope for: the simulated total interfering signal, a sinusoid and an "impulsive" signal, in the time domain.

To study how long it takes from the moment the envelope exceeds a threshold to when it goes above it again, we use the Average Cross Duration (ACD). For each threshold, the ACD is the Average of the Upgoing Crossing Times. Fig. 10 shows a representation of the normalized ACD for the three signals discussed in the previous section. In Fig. 10 the first threshold value corresponding to 0 $\mu$V is not represented, for better viewing of the graph. However, its value, which coincides with the maximum value of the ACD, is displayed in the legend of Fig. 10. Therefore, these minimum values of envelope are present much longer than the rest of envelope values. As in the case of the LCD, the maximum value of the ACD and their distribution are different from those corresponding to the sinusoid and to the "impulsive" signal.

In Fig. 10, the abrupt changes in value of the ACD to certain thresholds are due to the superposition of individual interferences with very different values for the durations of the level crossings (in this case, most with small durations). Thus, in the case of CW-Morse interferences, it is an indication that various individual interferences exist simultaneously with very different transmission speeds, i.e., with very different values of WPM.

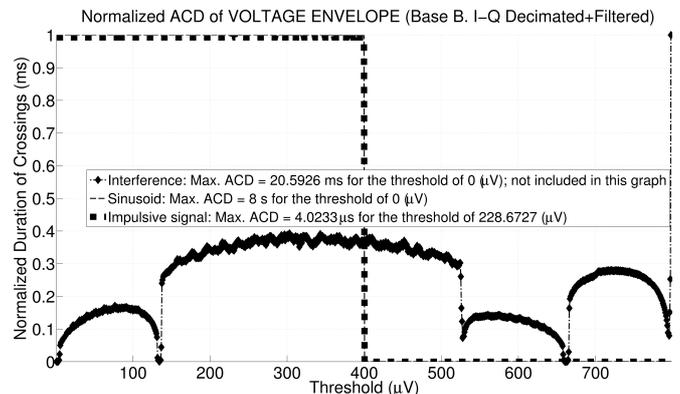

Fig. 10. Normalized ACD of the voltage envelope for: the simulated total interfering signal, a sinusoid and an "impulsive" signal, in the time domain.

Fig. 11 shows the LCD of the power envelope in the time domain. The aim is to compare with the pdf of the individual interference powers, shown in Fig. 3. In Fig. 11 can be noticed that the most frequent value is -87.3495 dBm, being -107.3391 dBm the expected value for the pdf of Fig. 3. This higher power value is to be expected since the LCD of Fig. 11 is for the total interfering signal, while the values in Fig. 3 are the average power of each individual interference during its lifetime.

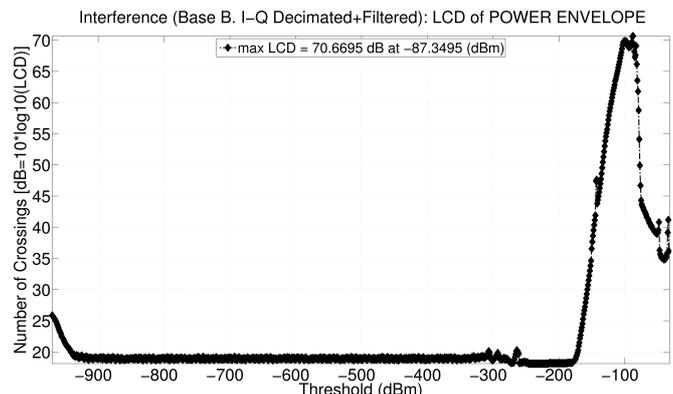

Fig. 11. The LCD of the power envelope for the simulated total signal.

## VIII. FUTURE WORK

It is evident that the validity of the implemented simulator must be proved by comparing the results of the simulation with real interference measurements. The tests of the model should include amateur radio, using: CW-Morse, SSB and even RTTY (Radio TeleTYpe). The tests should also include other kind of common users in the HF band, such as AM stations and the military HF radio stations (using, for example, M-ary

PSK). There is a high probability that PLT (Power Line Telecommunications) would cause increased noise levels at sensitive receiver sites given the existing and projected market penetration [23]. Therefore, the tests should also include possible interference of PLC (Power Line Communications) systems (using, for example, Orthogonal Frequency Division Multiplexing).

Future work should include congestion models including updated forecasts based on time of day, beyond our weekly model that was focused on noon and midnight. Also, it must be remembered that the model of congestion has been developed on measurements taken in Northern Europe. It would be desirable for the models of congestion also to include long range communications using near vertical incidence (NVIS), i.e. monopole high angles.

A possible application of the interference simulator beyond HF may be in the Global Positioning System (GPS) band. As in the case of HF, the simulator would be a tool for evaluating communication systems (GPS in this case) as well as potential new methods to combat interference.

## IX. CONCLUSIONS

An interference simulator for the whole HF band (1,606 – 30 MHz) has been modeled and implemented. The simulator generate interfering signals that can be found in a given frequency allocation, in a given time (past, present or future) and for a given location. The simulator has been used to simulate CW-Morse interference of the "ARRL Field Day 2011". It have highlighted some statistical measures useful in the analysis of interferences, especially in the case of CW-Morse interferences.

Our most important original contributions are:

1) To create a new model and simulator, we used jointly and in detail two existing independent models: a congestion model and a model of narrowband interference. Basically, the user of the simulator selects the date and location as well as the modulation assigned to each frequency allocation. So that our model, to generate the simulated signal for all the HF frequency allocations, provide at each instant, the number of interferences, their frequencies, their amplitudes, their temporal onset, duration, etc.

2) Our model does not require measurements.

3) As a result, our simulator is characterized by its ease of use and the freedom it offers to choose scene (modulation, location, week, year, etc.).

4) In addition, we have defined a generic modulating function and the conditions to model a "contact" CW-Morse, who meets the usual standards of contest.

5) Consequently, our interference model in conjunction with the CW-Morse modulating function designed, it results in a specific model for CW-Morse amateur contests. The major radio amateur contests can be a good scene to evaluate a communications system subjected to a high degree of "interferences".

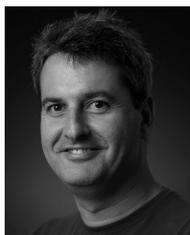

**Eduardo Mendieta Otero** was born in La Habana, Cuba, in 1966. He received B.Sc. (Hons.) degree in Electronic Equipment and B.Sc. (Hons.) degree in Radio Communication from University of Las Palmas de Gran Canaria (ULPGC), Spain, in 1992 and 1994. He received the M.S. degree in Telecom Engineering from ULPGC, Spain, in 2001.

He joined the ULPGC (Spain) as an Associate Professor in 1994. From 2006 to 2010, he was a member of the Technological Centre for Innovations in Communications (CeTIC). From 2010, he is member of Institute for Technological Development and Innovation in Communications (IDeTIC), Spain. His research activities are mainly in the field of Signal Processing for Radio Communications.

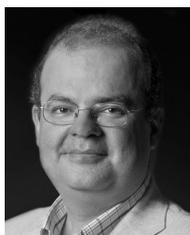

**Iván A. Pérez Álvarez** (M'89) was born in Gran Canaria, Spain, in 1965. Ivan A. Pérez-Álvarez is Telecom Engineer (M.S. degree) and Dr. Engineer by the Universidad Politécnica de Madrid (Spain) in 1990 and 2000, respectively.

From 1989 to 1997 he was in Europea de Comunicaciones S.A. and Telefónica Sistemas S.A., where he was a Member Staff of Special Projects Department and working in the Digital Signal Processing Group. He was involved in the design of HF/VHF/UHF digital communications systems for the Spanish Ministry of Defence. He joined the ULPGC (Spain) as an Associate Professor in 1998. At 2006 he was Head of CeTIC until 2010 and at the present is the Vice-Director of IDeTIC (Spain). His research activities are mainly in the field of Signal Processing for Radio Communications.

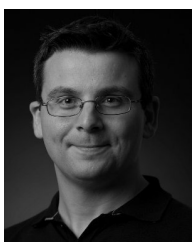

**Baltasar Pérez Díaz** was born in Tenerife, Spain, in 1975. Baltasar Pérez Díaz is a Telecom Engineer (M.S. degree) by ULPGC, Spain, in 2005.

He is currently working as a Research Assistant for IDeTIC (Spain) in hardware developments on HF broadband radiocommunications. His research activities are focused on broadband radiocommunications, RF subsystems, coupled-oscillator arrays on microwave bands and millimetre band radar.